\documentclass{aa}
\usepackage{txfonts}
\usepackage{graphicx}
\usepackage{natbib}

\bibpunct{(}{)}{;}{a}{}{,} 

\newcommand{\gr}{$\gamma$-ray \,}
\newcommand{\grs}{$\gamma$-rays \,}
\begin{document}
  \title{Dependence of the gamma-ray emission from SN~1006 on the
astronomical parameters}

\titlerunning{Gamma-ray emission from SN~1006}

  \author{L.T.Ksenofontov
          \inst{1}
          \and
           E.G.Berezhko
          \inst{1}
          \and
          H.J.V\"olk
          \inst{2}}

   \offprints{H.J.V\"olk}

   \institute{Yu.G.Shafer Institute of Cosmophysical Research and Aeronomy,
                     31 Lenin Ave., 677980 Yakutsk, Russia\\
              \email{berezhko@ikfia.ysn.ru}
              \email{ksenofon@ikfia.ysn.ru}
         \and
             Max-Planck-Institut f\"ur Kernphysik,
                Postfach 103980, D-69029 Heidelberg, Germany\\
             \email{Heinrich.Voelk@mpi-hd.mpg.de}
             }

   \date{Received month day, year; accepted month day, year}

\abstract
{We use nonlinear kinetic theory to study
the remnant dynamics and the particle acceleration as well as the 
properties of the nonthermal emission from
the supernova remnant SN~1006. The known range of astronomical
parameters is examined to determine whether it 
encompasses the
existing synchrotron emission data. Given the present-day spatial extent
and expansion rate of the object, it is shown that the hadronic
gamma-ray flux is very sensitive to the ambient gas density $N_{\mathrm
H}$ and that the existing H.E.S.S. upper limit requires $N_{\mathrm
H} < 0.1$~cm$^{-3}$. The strength of the amplified magnetic field downstream 
of the
shock is about 150~$\mu$G.  

}

\maketitle

%

\section{Introduction}

Despite the widespread belief that the cosmic rays (CRs) below the so-called 
knee in the energy spectrum at a few $10^{15}$~eV are accelerated in the 
shell-type supernova remnants (SNRs) of our Galaxy, this proposition  
still has
only limited evidence.  
This is not due to an inadequate theoretical understanding. 
In fact, essentially all physical processes involved have either been clarified 
or, like the injection rates of ions and of electrons into the diffusive 
shock acceleration process, can be inferred from the observed radio and X-ray 
synchrotron spectra \citep[for a recent overview, see][]{voelk04}. The 
limitation is rather the scarcity of SNR detections in TeV \grs as an 
unequivocal signature of the existence of very high energy particles in these 
sources.

Of this SN~1006 (G327.6+14.6) is the most prominent case. 
Observations of nonthermal X-rays \citep{koyama95} suggest that at least CR 
electrons are accelerated in SN~1006 up to energies of about 100~TeV. 
Subsequent \gr observations with the CANGAROO telescopes 
\citep{tanimori98,tanimori01} strengthened this conclusion. It was also shown 
\citep{bkv02} that nonlinear kinetic theory of CR acceleration is consistent 
with all observational data for a value $N_\mathrm{H}=0.3$~cm$^{-3}$ of the 
ambient interstellar medium (ISM) density from the range $0.05 \leq 
N_\mathrm{H} \leq 0.3$~cm$^{-3}$ existing in the literature. However, SN~1006 
could not be detected by the H.E.S.S. experiment as a TeV source in a total of 
18.2h (in 2003) and 6.3h (in 2004) {\it livetime} of ON source observations with 
SN~1006 in the field of view \citep{aharonian05}. The H.E.S.S. upper limit is 
roughly one order of magnitude lower than the published 
CANGAROO flux.

In this paper we demonstrate that the H.E.S.S. upper limit does
not invalidate the theoretical picture \citep{bkv02} on which the
previous calculation of the \gr emission spectrum has been based. As we
shall show, it is rather the value of the external astronomical
parameter $N_\mathrm{H}$ that strongly influences the hadronic \gr
flux. The Inverse Compton (IC) \gr flux, on the other hand, depends only
weakly on $N_\mathrm{H}$. We use this theory to 
describe all the
relevant properties of SN~1006 and re-examine the most relevant
set of physical and astronomical parameters, mainly $N_\mathrm{H}$, the
total explosion energy and, to a minor extent, the interior magnetic
field strength as well as the distance to the object, given the H.E.S.S.
upper limit.

\section{SNR parameters}

Since SN~1006 is a type Ia supernova it presumably expands into a uniform ISM, 
ejecting roughly a Chandrasekhar mass $M_\mathrm{ej}=1.4 M_{\odot}$, which is 
characterized by the initial velocity distribution $dM_\mathrm{ej}/dv\propto 
v^{2-k}$ with $k=7$. We note that, since SN~1006 is already in the Sedov phase, 
its properties are not sensitive to the parameter values 
$M_\mathrm{ej}$ and 
$k$ of the ejecta. The ISM gas density $\rho_0=1.4m_\mathrm{p}N_\mathrm{H}$, 
which is usually characterized by the hydrogen number density $N_\mathrm{H}$, is 
an important parameter that strongly influences the expected 
SNR dynamics and 
nonthermal emission. However, the value of $N_\mathrm{H}$ is poorly constrained 
for SN~1006 from the thermal emission observations. They yield a wide range of 
possible values from $N_\mathrm{H}=0.05-0.1$~cm$^{-3}$ \citep{dwarkadas98} to 
$N_{\mathrm H}\approx 0.3$~cm$^{-3}$ \citep{dubner02}.

As in our early study \citep{bkv02} we apply here nonlinear kinetic theory
of CR acceleration in SNRs \citep{bek96,bv97} to find the optimum set of
physical parameters of SN~1006, mainly the value of $N_\mathrm{H}$, which
give a consistent description of the observed overall dynamics and of the
nonthermal emission. As described in detail in these papers, that theory
includes all the important physical factors which influence CR
acceleration and SNR dynamics: shock modification by the CR backreaction,
Alfv\'en wave damping within the shock transition, selfconsistent
determination of the CR spectrum, and spatial distribution in each
evolutionary phase. In addition it includes synchrotron losses of CR
electrons and a determination of all nonthermal emission processes 
produced in SNRs by accelerated CRs \citep[e.g.][]{bkv02,bpv03a}. It was
demonstrated for the cases of SN~1006 \citep{bkv02} and Cassiopeia~A
\citep{bpv03a} that the values of these key parameters (proton injection
rate, electron to proton ratio and interior magnetic field strength) which
cannot be predicted theoretically with the required accuracy, can be
determined from a fit of the observed synchrotron emission data. It is 
important here that the parameter values for these SNRs, determined
in this way, were very well confirmed by the Chandra measurements of the
fine structure of the nonthermal X-ray emission \citep{bkv03b,bv04}.

We take as the most reliable estimate for the distance $d=2.2$~kpc
to SN~1006 \citep{winkler03}. This distance is roughly 20\% larger than
the value $d=1.8$~kpc adopted in \citet{bkv02}.

\subsection{Ambient gas density and total explosion energy}

%
\begin{figure}[t]
 \includegraphics[width=.47\textwidth]{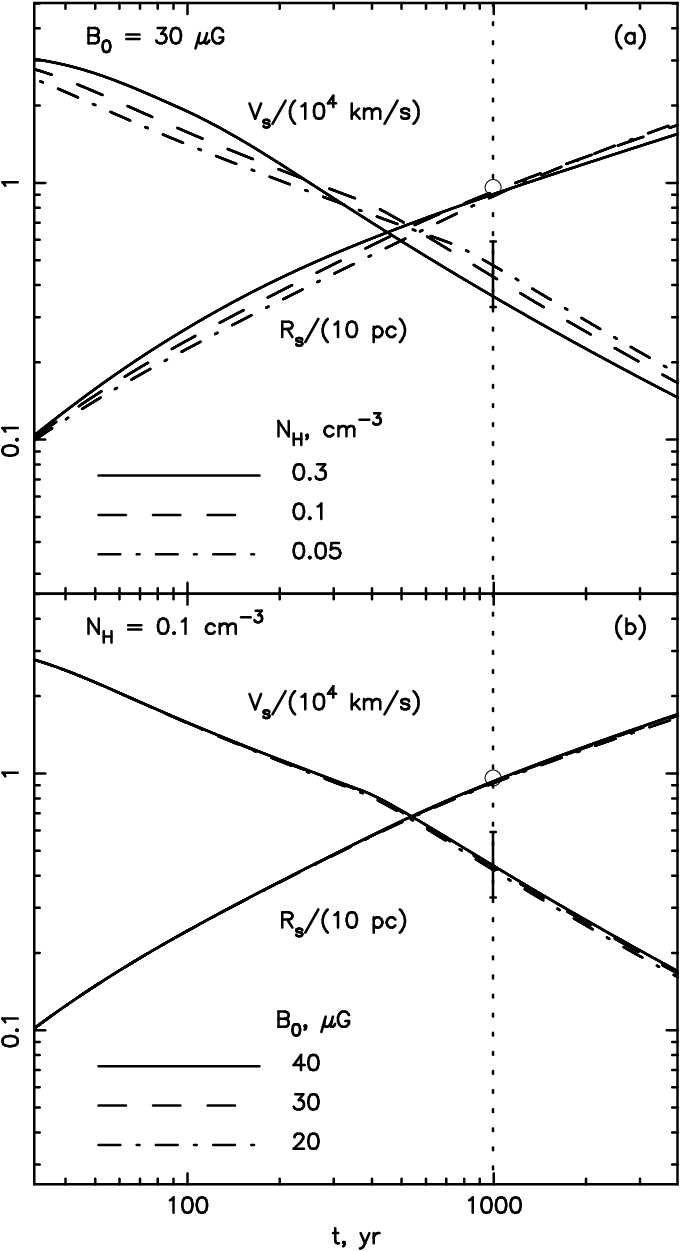}
 \caption{Shock radius $R_\mathrm{s}$ and shock speed $V_\mathrm{s}$
as a function of time for the upstream magnetic field $B_0=30$~$\mu$G and
different ISM number densities $N_\mathrm{H}$ (a), and for ISM number
density $N_\mathrm{H}=0.1$ cm$^{-3}$ and different values of the upstream
magnetic field $B_0$ (b). The observed size and speed of the shock
\citep{moffett93} are shown as well.}
\label{fig1}
\end{figure}

The SNR dynamics are illustrated in Fig.~\ref{fig1}. In 
Fig.~\ref{fig1}a
we present the time dependence of the shock radius $R_\mathrm{s}$ and
of the shock speed $V_\mathrm{s}$, calculated for the optimum upstream
magnetic field value $B_0=30$~$\mu$G (see below), and for the expected
density range of the ambient ISM. The value of the explosion energy was
taken to fit the observed size and speed at the current epoch $t\approx
10^3$~yr. Since SN~1006 has already evolved into the Sedov phase, the
explosion energy scales as $E_\mathrm{sn}\propto N_\mathrm{H}$. The
calculations correspond to $E_\mathrm{sn}/(10^{51}$~erg) = 1.9, 3.8 and
11.4, for $N_\mathrm{H}/(1~\mbox{cm}^{-3})$= 0.05, 0.1 and 0.3,
respectively.

The scaling $E_\mathrm{sn}\propto N_\mathrm{H}$ is intuitively plausible, since 
for lower external density both the expansion velocity as well as the radius 
$R_\mathrm{s}(t)$ of the SNR are larger for a given explosion energy 
$E_\mathrm{sn}$. Asymptotically -- and 
approximately, since there exists no strictly self-similar solution with 
particle acceleration --  in the Sedov phase, we have $R_\mathrm{s}(t)\propto 
(E_\mathrm{sn}/N_\mathrm{H})^{1/5}t^{2/5}$.  Therefore the ratio 
$E_\mathrm{sn}/N_\mathrm{H}$ is fixed for known $t$, $R_\mathrm{s}(t)$ and 
$V_\mathrm{s}(t) = dR_s(t)/dt$.

\begin{figure}[t]
\includegraphics[width=.47\textwidth]{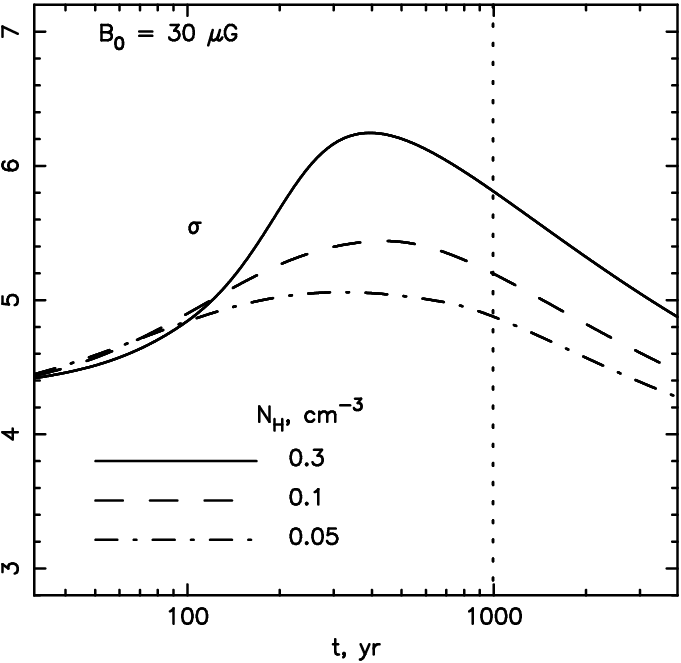}
 \caption{The overall shock compression ratio $\sigma$ as a function of
time, for three different ISM densities $N_\mathrm{H}=0.05$, 0.1, and
0.3~cm$^{-3}$, and upstream field $B_0=30$~$\mu$G.}
\label{fig2}
\end{figure}

The calculations shown in Fig.~\ref{fig1}a demonstrate that with the
dependence of $E_\mathrm{sn}$~on $N_\mathrm{H}$~the SNR dynamics are indeed
compatible with the measured values of $R_\mathrm{s}(t)$ and
$V_\mathrm{s}(t)$ for different ISM densities. We note that the small
differences in $V_\mathrm{s}(t)$ for different $N_\mathrm{H}$ in Fig.1a
during the Sedov phase $t>400$~yr are the result of the shock
modification, which depends on $N_\mathrm{H}$ (see Fig.~\ref{fig2}).

In addition, Fig.~\ref{fig1}b demonstrates the expected weak dependence of
the dynamics on the magnetic field strength $B_0$. For a higher magnetic
field $B_0$ the shock compression ratio $\sigma$ is lower (compare
Fig.~\ref{fig2} with Fig. 1b in \citet{bkv02}).
Therefore the postshock internal pressure $P_2\approx \rho_0
V_\mathrm{s}^2(1-1/\sigma)$ is higher for lower $B_0$. Since 
in the Sedov
phase the thermal energy is the dominant form of energy, the total energy
is $E_\mathrm{sn}\propto R_\mathrm{s}^3P_2$. Since $E_\mathrm{sn}$ is a
constant the shock size $R_\mathrm{s}$ and therefore the shock speed
$V_\mathrm{s}$ are somewhat larger for larger magnetic field $B_0$, in
agreement with Fig.~\ref{fig1}.

\subsection{Amplified B-field}

In our model the interior (downstream) magnetic field
$B_\mathrm{d}$ is connected to the outer (upstream) magnetic 
field
$B_0$ by the simple relation $B_\mathrm{d}=\sigma B_0$, where $\sigma$
is the overall shock compression ratio \citep{bkv02}. It is also assumed
that already the upstream field $B_0$ is significantly amplified as a
consequence of CR streaming, and that it therefore substantially exceeds the 
existing
ISM value \citep{lb00,belll01,bell04}.

The shock compression ratio depends on the ambient gas density, as shown
in Fig.~\ref{fig2}.  The behavior of the shock compression ratio $\sigma$
is determined by the values of two parameters: the proton injection rate
$\eta$ and the Alfv$\acute{e}$nic shock Mach number
$M_\mathrm{a}=V_\mathrm{s}/c_\mathrm{a}$, where
$c_\mathrm{a}=B_0/\sqrt{4\pi \rho_0}$ is the Alfv$\acute{e}$n speed. The
value of the injection rate $\eta$ influences the critical shock speed
$V_\mathrm{s}^*\propto \eta$ \citep{berel99} which separates the initial
SNR evolutionary period, when $V_\mathrm{s}>V_\mathrm{s}^*$ and therefore
the shock is almost unmodified, from the subsequent period
$V_\mathrm{s}<V_\mathrm{s}^*$ of the modified shock.  In our case
$\eta=2\times 10^{-4}$ and $V_\mathrm{s}^*\approx 10^3$~km/s. During the
period $V_\mathrm{s}<V_\mathrm{s}^*$ the shock compression ratio goes as
$\sigma \approx 1.5 M_\mathrm{a}^{3/8}$ \citep{berel99}. Therefore a
decrease of the ISM density $N_\mathrm{H}$ leads to a
decrease of the shock compression ratio $\sigma\propto
V_\mathrm{s}^{3/8}
N_\mathrm{H}^{3/16}$, as seen in Fig.~\ref{fig2} for $t>400$~yr (see
also Fig.~\ref{fig1}a). Thus, for the
case $B_0=30$~$\mu$G used in Fig.~\ref{fig1}a, the downstream magnetic
field strengths are $B_\mathrm{d}= 149$, 156 and 173~$\mu$G for
$N_\mathrm{H}=0.05$, 0.1 and 0.3~cm$^{-3}$, respectively.

\subsubsection{Spatially integrated synchrotron spectrum}

\begin{figure}[t]
 \includegraphics[width=.47\textwidth]{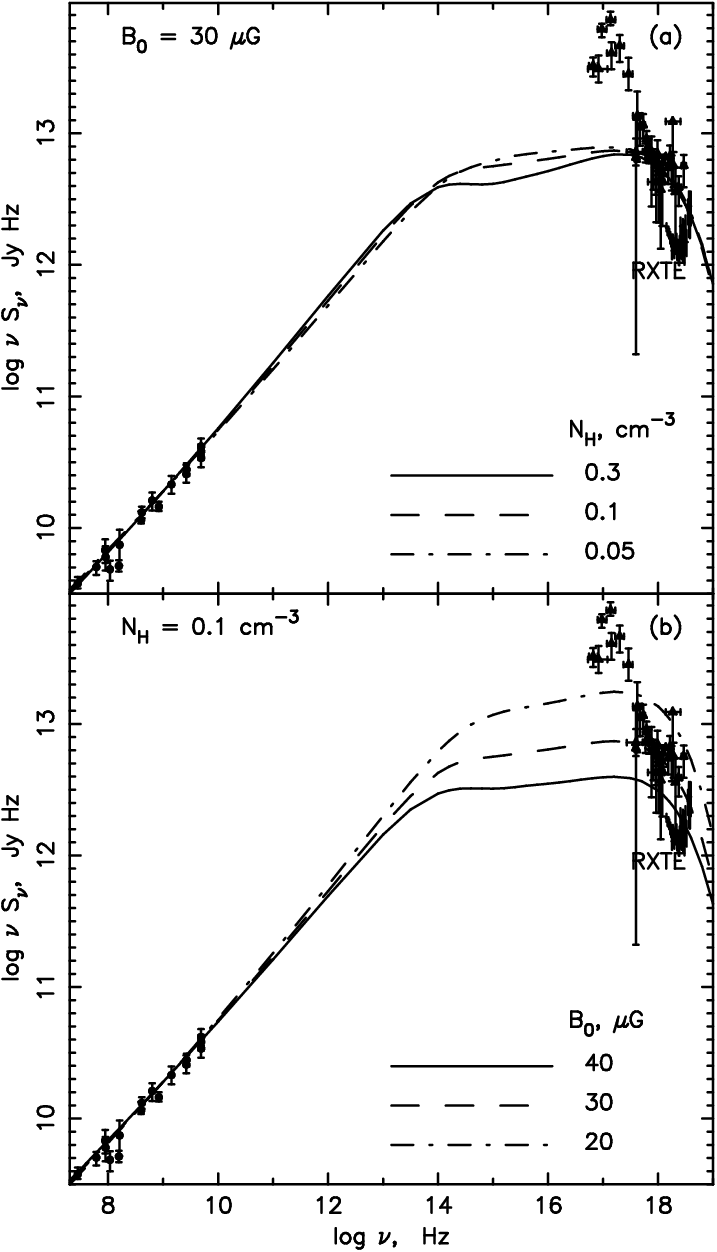}
 \caption{Synchrotron emission flux as a function of frequency for the
 same cases as in Fig.~\ref{fig1}. The observed X-ray
 \citep{Hamilton86,allen99} and radio emission \citep{reynolds96} 
  fluxes are shown.} 
\label{fig3}
\end{figure}

For the global determination of the effective downstream field
$B_\mathrm{d}$ we compare the theoretical synchrotron spectrum,
calculated for three different values of $B_0$, with the observed
spatially integrated spectrum (see Fig.~\ref{fig3}). 

Electrons with a power-law energy spectrum $N_e(\epsilon)= A_e \epsilon
^{-\gamma}$ produce a synchrotron flux $S_{\nu}=A\nu ^{-\alpha}$ with
spectral index $\alpha=(\gamma-1)/2$ and amplitude $A\propto A_e
B_\mathrm{d}^{(\gamma+1)/2}/d^2$. In the test-particle limit, 
$\gamma=2$ and therefore
$\alpha=0.5$. Values $\alpha>0.5$, as observed in young SNRs, require a
curved electron spectrum (hardening to higher energies) as predicted by
nonlinear shock acceleration models. The synchrotron emission at frequency
$\nu$ is mainly produced by electrons of energy $\epsilon=5\sqrt{
[\nu/(\mbox{1~GHz})] [(10~\mu\mbox{G})/B_\mathrm{d}] }$~GeV. Since the typical
particle spectrum produced by a CR modified shock \citep{bek96} is
characterized by $\gamma>2$ at $\epsilon<1$~GeV and $\gamma<2$ at
$\epsilon>10$~GeV it follows that in order to have $\alpha=0.57$, observed
for SN~1006, numerical iteration shows that one needs efficient CR acceleration with a proton injection
rate $\eta=2\times 10^{-4}$, which leads to the required shock
modification, and also to a high interior magnetic field $B_\mathrm{d}\ge
120$~$\mu$G \citep{bkv02}. X-ray synchrotron spectral measurements are
required to find the optimum value of the magnetic field strength
$B_\mathrm{d}$ \citep{bkv02,bpv03a}: for a given fit of the synchrotron
spectrum in the radio range the X-ray synchrotron amplitude is very
sensitive to $B_\mathrm{d}$ (see Fig.~\ref{fig3}b).

An increase of the ISM number density $N_\mathrm{H}$ leads to an increase of the 
total number of low energy CRs. Since the radio emission is produced by low 
energy electrons and since for known SNR angular size and expansion rate the 
shock radius $R_\mathrm{s}$ and shock speed $V_\mathrm{s}$ are proportional to 
distance $d$, we have $A_\mathrm{e}\propto \eta V_\mathrm{s}R_\mathrm{s}^3 
K_\mathrm{ep}N_\mathrm{H} \propto \eta d^4 K_\mathrm{ep}N_\mathrm{H}$. Therefore 
the amplitude of the synchrotron flux $A\propto \eta 
K_\mathrm{ep}N_\mathrm{H}d^2 B_\mathrm{d}^{(\gamma+1)/2}$. Since $A$ is fixed by 
the experiment, the electron to proton ratio changes like 
$K_\mathrm{ep} \propto 
N_\mathrm{H}^{-1}$ for a given injection rate $\eta$ and interior magnetic 
field 
$B_\mathrm{d}$. Therefore we have $K_\mathrm{ep} = 2.6\times 10^{-3}, 1.1\times 
10^{-3}$ and $4.2\times 10^{-4}$ for $N_\mathrm{H}=0.05,$ 0.1 and 0.3~cm$^{-3}$ 
respectively. This includes the renormalization factor 0.2 for the overall 
ion intensity. For a given $N_\mathrm{H}$ we have $K_\mathrm{ep} \propto d^{-2} 
B_\mathrm{d}^{-(\gamma+1)/2}$. Given that the Galactic CRs have a value 
$K_\mathrm{ep} \simeq 10^{-2}$, the small value of $N_H$ suggested by the H.E.S.S. 
experiment makes SN 1006 -- at the present stage of evolution -- more closely 
resemble an average source of the Galactic CRs.

As a result the synchrotron spectrum is almost insensitive to
$N_\mathrm{H}$. The weak dependence of $S_{\nu}(\nu)$ on $N_\mathrm{H}$ is
caused by the different downstream magnetic field $B_\mathrm{d}=\sigma
B_0$ at the same $B_0$, because the compression ratio $\sigma$ depends on
$N_\mathrm{H}$. For larger $N_\mathrm{H}$ the shock is more strongly
modified. This leads to a harder CR spectrum at high energies $\epsilon
\gg m_\mathrm{p}c^2$ and therefore the spectrum $S_{\nu}(\nu)$ at
$\nu=10^{15}-10^{18}$~Hz becomes harder for larger $N_\mathrm{H}$, as can
be seen in Fig.~\ref{fig3}a.

In addition, an increase of the magnetic field $B_\mathrm{d}$ leads to a
decrease of the electron energy $\epsilon_\mathrm{l}\propto
t^{-1}B_\mathrm{d}^{-2}$ \citep{bkv02}, where the electron spectrum
undergoes a brake: for $\epsilon> \epsilon_\mathrm{l}$ synchrotron losses
in the downstream region are significant and lead to a steep overall
(spatially integrated) electron spectrum $N_\mathrm{e}\propto \epsilon
^{-\gamma-1}$, where $\gamma$ is the power law index of the CR spectrum at
the shock front. Therefore the synchrotron spectrum has a break at the
corresponding frequency $\nu_\mathrm{l}\propto
\epsilon_\mathrm{l}^2B_\mathrm{d} \propto B_\mathrm{d}^{-3}$, which
decreases with increasing $B_\mathrm{d}$, as one can see in
Fig.~\ref{fig3}.

An increase of the magnetic field requires a reduction of the amplitude of the 
electron momentum distribution since the radio electrons are not subject to 
synchrotron cooling at the present epoch. This leads to a decreased 
electron/proton ratio $K_\mathrm{ep}\propto B_0^{-\alpha-1}$ in the energy 
range where radiative cooling is unimportant. Most importantly however, the 
cooling region of the synchrotron emission spectrum $\nu>\nu_\mathrm{l}$ will be 
lowered in amplitude by this decreasing electron/proton ratio. Comparison with 
the experimental X-ray data shows that the optimum magnetic field value is about 
$B_0=30$~$\mu$G, and corresponds to a downstream field $B_{\mathrm d}\approx 
150$~$\mu$G. This is somewhat larger than the $B_{\mathrm d}\approx 120$~$\mu$G 
quite conservatively estimated earlier \citep{bkv02,bkv03b}, and is in 
agreement with the field amplification that is implied by the filamentary 
structures in hard X-rays, observed with Chandra \citep{bamba03,vbk04} (see also 
below).

As it was already noted (Berezhko et al. 2002), it is only the
highest energy part of the X-ray spectrum corresponding to $\nu
>10^{18}$~Hz that is of nonthermal origin. At lower frequencies $\nu \ll
10^{18}$~Hz the spectrum is dominated by thermal emission. Therefore the
theoretical spectrum for $\nu \ll 10^{18}$~Hz should be substantially
below the experimental values, as it is in Fig.3.

\subsubsection{Overall synchrotron morphology}
\begin{figure}[t]
 \includegraphics[width=.47\textwidth]{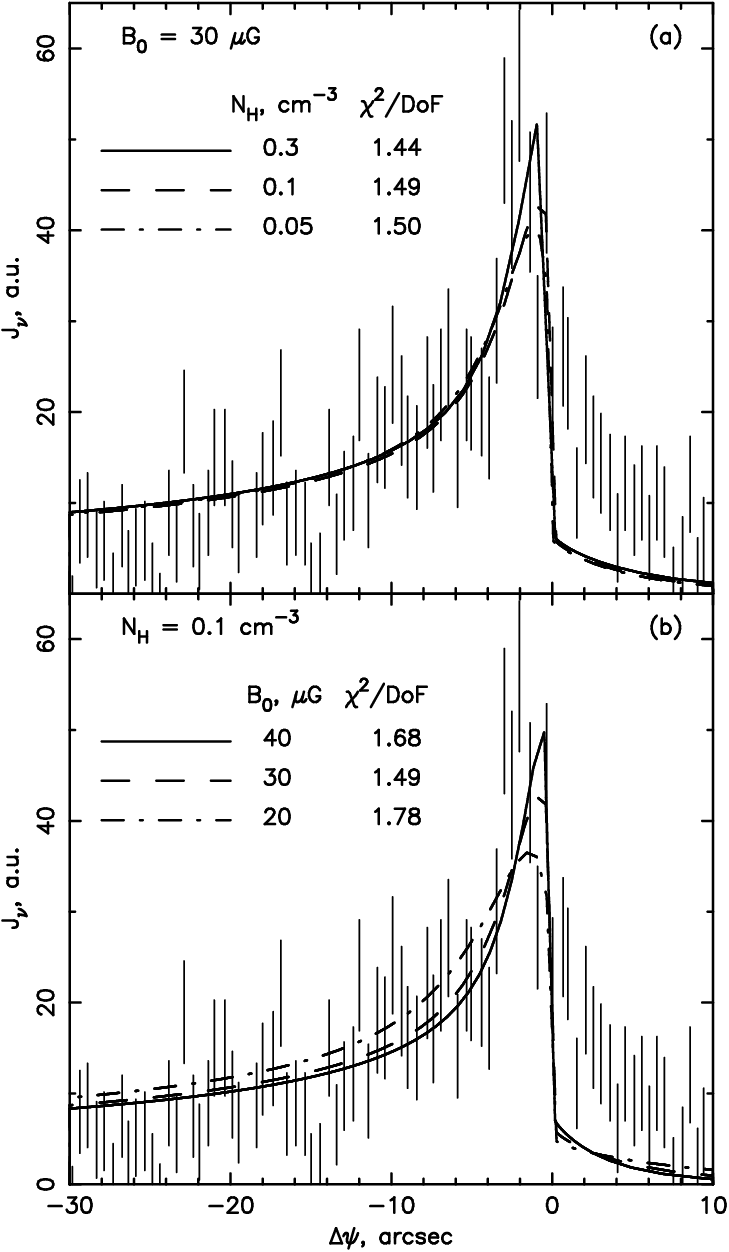}
 \caption{Projected radial dependence of the non-thermal X-ray 
 brightness in the energy range 2 to 10 keV, calculated for the same 
 cases as in Fig.~\ref{fig1} together with the Chandra data 
  corresponding to the sharpest profile in 
 \citep{bamba03}.}
\label{fig4}
\end{figure}

We now compare this globally determined magnetic field value $B_0=30$~$\mu$G 
with the local Chandra results by plotting the full numerical solutions for the 
overall SNR morphology, as it results from selfconsistent time-dependent kinetic 
theory, jointly with the sharpest of the experimental brightness profiles 
obtained by the Chandra observers \citep{bamba03} in Fig.~\ref{fig4}a, where the 
values of $\chi^2/dof$ characterizing the quality of fit are also shown. Since 
this profile obviously lies on a high superimposed background, 
we have 
subtracted from the data the constant brightness value which corresponds to the 
far upstream region. Note that the initial Chandra data were presented as a 
function of the angular radial distance $\psi$ \citep{bamba03}. The 
transformation to the linear projection distance $\rho=\psi d$ has been achieved 
by assuming a distance of $d=2.2$~kpc to SN~1006 (see above). We use in 
Fig.~\ref{fig4},~\ref{fig5} $\Delta \psi$ instead of $\psi$, taking the initial 
point $\Delta \psi=0$ at the theoretically predicted shock position. The 
agreement of global morphology and local profiles is reasonable for all 
considered ISM densities.

Since the absolute values of the experimental brightness profiles 
and the actual shock position are not known their values were used as adjusting
parameters in the fitting procedure.

Note that the radial profiles shown in Fig.~\ref{fig4}a depend on $N_\mathrm{H}$ as a 
result of the dependence of the downstream magnetic field $B_\mathrm{d}=\sigma 
B_0$ on $N_\mathrm{H}$: for higher $N_\mathrm{H}$ the compression ratio $\sigma$ 
is larger (see Fig.2) and leads to a higher $B_\mathrm{d}$ and therefore to a 
thinner profile. In addition, the radial profile depends on the downstream 
plasma speed $u_2=V_\mathrm{s}/\sigma$ (Berezhko \& V\"olk 2004). The lower 
value of $u_2$, that is associated with a larger $N_\mathrm{H}$, makes the 
profile thinner.

As can be seen from Fig.~\ref{fig4}b, the magnetic field value
$B_0=20$~$\mu$G slightly overestimates the profile width. Together with
the spectral fit (Fig.~\ref{fig3}b) we conclude that $B_0=30$~$\mu$G
($B_\mathrm{d}\approx 150$~$\mu$G) is a realistic estimate.

\subsubsection{Local fit of the synchrotron profile} %
\begin{figure}[t]
 \includegraphics[width=.47\textwidth]{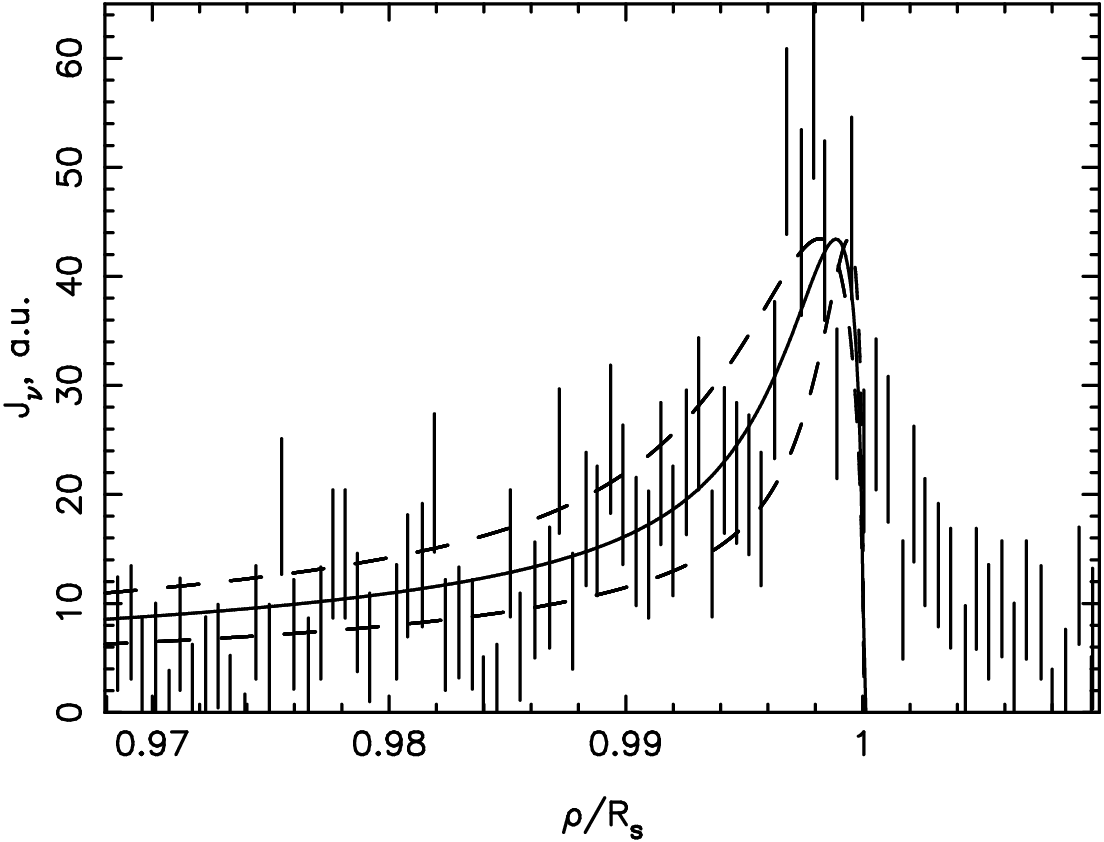}
\caption{The same Chandra 2-10~keV radial continuum brightness profile
as in Fig.~\ref{fig4}, fitted to the projection of the
exponential emissivity profile described by the model of Eq.~(\ref{eq2}).
The fit has a $\chi^2/dof=1.2$. The dashed lines indicate the $1\sigma$
deviation.}
\label{fig5}
\end{figure}

X-ray brightness profiles 
can also be used for an independent determination of the interior magnetic 
field value that does not require the nonlinear theory 
(Fig.~\ref{fig5}).

It has been shown \citep{bkv03b,bv04} that in the case of strong synchrotron 
losses, which in fact occur for the electrons responsible for the X-ray 
synchrotron radiation, the {\it downstream diffusion length over a synchrotron 
loss time} mainly determines the three-dimensional thickness of the X-ray 
emission profile. The effective magnetic field can then be expressed in terms of 
the width $l_2$ of the radial synchrotron emissivity profile $q_{\nu}(r)$ 
according to the simple relation
\begin{equation}
B_\mathrm{d}=\left({3m_\mathrm{e}^2c^4\over 4er_0^2l_2^2}\right)^{1/3}
\approx 0.5 \left({10^{16}\mbox{ cm}\over l_2}\right)^{2/3} \mbox{ mG},
\label{eq1}
\end{equation}
where $m_\mathrm{e}$, $e$ and $r_0$ denote the electron mass, charge, and
classical radius, respectively. The actual value of $l_2 $ 
is determined from the observed width of the projected radial
brightness profile
\[ J_{\nu}(\rho)=\int_{-\infty}^{\infty}q_{\nu}(r=\sqrt{\rho^2+x^2})dx,
\]
where $\rho$ is the projected distance from the remnant center and the
integration is performed along the line of sight.

Since the synchrotron emissivity can be represented in the form $q_\nu(r)= q_2 
\exp [(r-R_\mathrm{s})/l_2]$ \citep{bkv02}, the projected profile is expressed 
in the analytical form
\begin{equation}
J_{\nu} \approx q_2 {R_\mathrm{s}\sqrt{2\pi R_\mathrm{s} l_2} \over \rho} 
\exp(-t^2) \mbox{ erfi } (t),
\label{eq2}
\end{equation}
where $t=R_\mathrm{s} \sqrt{R_\mathrm{s}^2-\rho^2}/ (\rho \sqrt{2
R_\mathrm{s} l_2})$, and where erfi($x$) is an imaginary error
function. Note that the profile $J_{\nu}(\rho)$ can be expressed in
terms of elementary functions \citep{bv04}, if we use the expansion of
erfi$(t)$ near the point $t=0$.

As shown in Fig.~\ref{fig5}, the experimental Chandra values are quite
well fitted by the analytical profile, described by Eq.~(\ref{eq2}), with
an emissivity width $l_2=13.7 (+8.2-6.2)\times 10^{-3}$~pc, that results
in the effective magnetic field $B_\mathrm{d}=191 (+82-62)$~$\mu$G. This
range of downstream B-field strengths includes the values following for
the three densities considered in Fig.~\ref{fig1}a on the basis of an
upstream field of $B_0=30$~$\mu$G which we shall use for the following.

Taking into account these results together with the consideration
presented in Fig.3 we conclude that $B_0=30$~$\mu$G and
$B_\mathrm{d}\approx 150$~$\mu$G are the most appropriate values
consistent with the existing data.

\section{Nonthermal energy density}

\begin{figure}[t]
 \includegraphics[width=.47\textwidth]{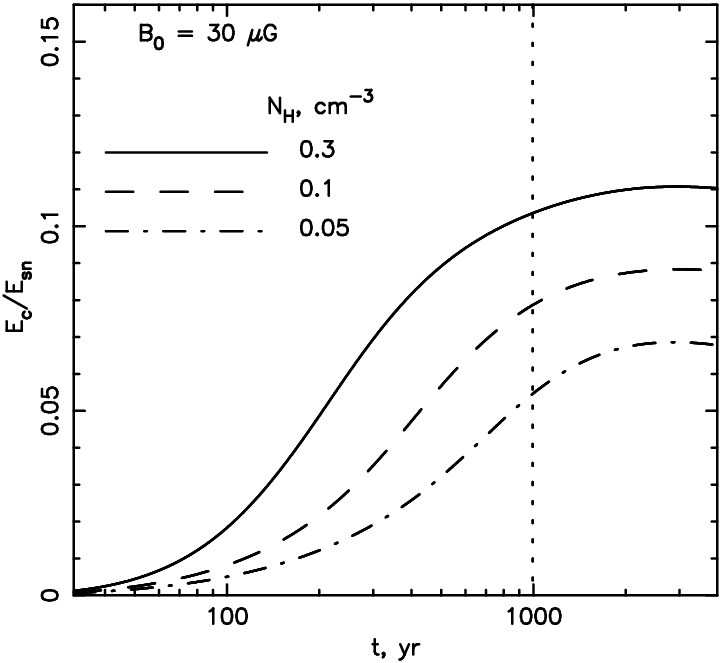}
\caption{Total energy $E_\mathrm{c}/E_\mathrm{sn}$ of accelerated
particles in the SNR, normalized to the total hydrodynamic explosion
energy $E_\mathrm{sn}$, as a function of time for the different ISM
densities $N_\mathrm{H}$ shown in Figs. 1-5, assuming spherical
symmetry. $E_\mathrm{c}$ is renormalized by a factor
of 0.2 \citep[cf.][]{vbk03}. The upstream magnetic field strength is
$B_0=30$~$\mu$~G.}
\label{fig6}
\end{figure}

As we have seen, $E_\mathrm{sn}\propto N_\mathrm{H}$ in the Sedov phase, for 
given SNR size and expansion rate. Because these parameters are independent of 
time, it is necessary to choose this proportionality generally. Fig.~\ref{fig1} 
demonstrates that this approach is roughly consistent for all phases. The 
resulting values for $E_{\mathrm c}$ in Fig.~\ref{fig6} are also calculated on 
this assumption. 

In Fig.~\ref{fig6} we present the relative CR energy content
$E_\mathrm{c}/E_\mathrm{sn}$ as a function of time, calculated for
$B_0=30$~$\mu$G and three different ISM number densities $N_\mathrm{H}$.
It increases with time, reaches a peak value at the epoch $t \simeq
3000$~yr and then slowly decreases due to an increasing dominance of
adiabatic cooling. In the early free expansion phase $t \ll 100$~yr, when
nonlinear shock modification is not essential (see Fig.~\ref{fig2}), the
shock produces a power law CR spectrum $f\propto N_\mathrm{inj}
p_\mathrm{inj} p^{-4}$, where $N_\mathrm{inj} =\eta N_\mathrm{H}$ is the
number of gas particles injected into the acceleration from each unit of
volume intersecting the shock front and $p_{\mathrm inj} \propto
V_\mathrm{s}$ is their momentum. Since the strong unmodified shock always
produces a CR spectrum of the same shape, the CR energy content
$E_\mathrm{c} \propto N_\mathrm{c}$ at this stage grows proportionally to
the total number $N_\mathrm{c}$ of accelerated CRs if we neglect weakly
time dependent factor $\ln(p_\mathrm{max}/m_pc)$, where $ p_\mathrm{max}$
is maximum CR momentum. Assuming spherical symmetry for the moment, and
starting from the obvious relation
\[ dN_\mathrm{c} \propto N_\mathrm{H} R_\mathrm{s}^2 V_\mathrm{s}^2 dt \]
and taking into account the shock expansion law \citep[eg.][]{chev82}
\[ R_\mathrm{s} \propto (E_\mathrm{sn}^2/N_\mathrm{H})^{1/7} t^{4/7} \]
we have
\begin{equation} 
E_\mathrm{c} \propto E_\mathrm{sn}^{8/7} N_\mathrm{H}^{3/7} t^{9/7}. 
\end{equation}
Then we have a residual density dependence at this early phase 
$E_\mathrm{c}/E_\mathrm{sn} \propto N_\mathrm{H}^{4/7}$
if we use $E_\mathrm{sn}\propto N_\mathrm{H}$.

In the transition to the Sedov phase (which happens earlier for larger
$N_\mathrm{H}$), $E_\mathrm{c}/E_\mathrm{sn}$ slowly reaches a maximum as
a function of time which is roughly independent of $N_\mathrm{H}$. This is
the result of the nonlinear limitation by the total amount of energy
$E_\mathrm{sn}$. During the Sedov phase the shock speed $V_{\mathrm s}(t)$
depends only on 
$E_\mathrm{sn}/N_\mathrm{H}$, as we have
seen before (see also Fig.~\ref{fig1}a), whereas the Alfv$\acute{e}$nic
shock Mach number $M_\mathrm{a} = V_\mathrm{s} / c_\mathrm{a} \propto
\sqrt{N_\mathrm{H}}$ increases with $N_{\mathrm H}$, because the
Alfv$\acute{e}$n speed $ c_\mathrm{a} \propto 1/\sqrt{N_{\mathrm H}}$.
This leads to a stronger shock modification (see Fig.~\ref{fig2}), that in
turn provides a more efficient CR production $E_\mathrm{c}/E_\mathrm{sn}$
for larger $N_\mathrm{H}$. In spherical symmetry $E_\mathrm{c}$ can reach
more than 50 \% of $E_\mathrm{sn}$ in this way. In reality, however, this
spherical symmetry does not exist, because for a SN explosion in a uniform
medium, as for SN~1006, the external magnetic field will also be 
uniform.
This leads to a severe reduction of the injection rate around those parts
of the shock surface that are roughly parallel to the external 
field and
requires a renormalization of the overall acceleration efficiency by a
factor of the order of $0.2$ \citep{vbk03}. This dipolar picture of the
synchrotron emission has been confirmed in an elegant
analysis of recent
{\it XMM} data for SN~1006 by \citet{roth04}. Altogether this leads to a
rough proportionality $E_\mathrm{c} \propto N_\mathrm{H}$ until particles
start to leave the remnant. As one can see from Fig.~\ref{fig6}, the
normalization of $E_\mathrm{c}$ by $E_\mathrm{sn}$ removes the main
density dependence: the relative difference
$E_\mathrm{c}(N_\mathrm{H}=0.3)/E_\mathrm{c}(N_\mathrm{H}=0.05) - 1$ of
the $E_\mathrm{c}$~values equals about 11, whereas the relative difference
of the normalized values $E_\mathrm{c}/E_\mathrm{sn}$ is about unity.

\begin{figure}[t]
 \includegraphics[width=.47\textwidth]{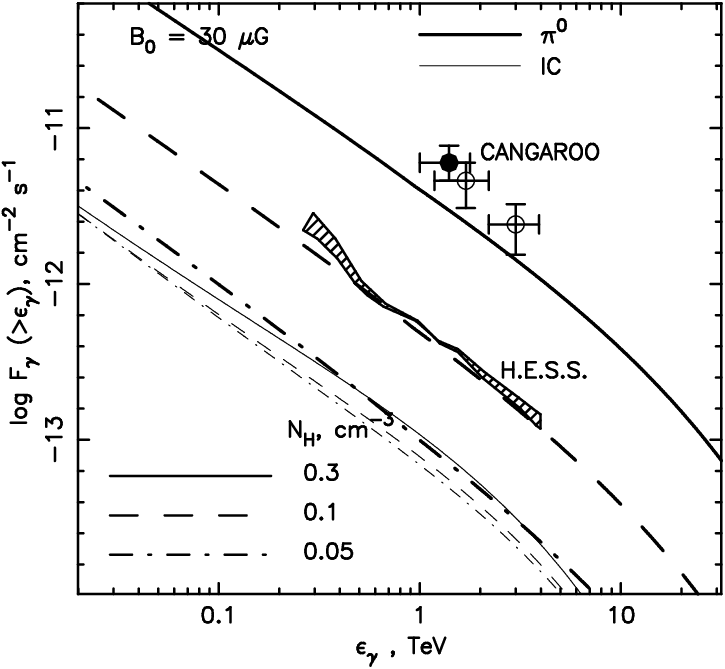}
 \caption{Integral $\pi^0$-decay
 ({\it thick lines}) and IC ({\it thin lines})
 $\gamma$-ray fluxes from the NE half of the remnant
 as a function of $\gamma$-ray energy for $B_0 = 30 \mu$G, and for the same
 different densities as in Fig.~\ref{fig1}. The CANGAROO-I ({\it open circles 
with 
error bars}) and CANGAROO-II ({\it solid circle with error bars}) flux from the 
NE rim
 \citep{hara02} and the corresponding H.E.S.S. upper
limit for the CANGAROO position
\citep{aharonian05} are shown as well. }
\label{fig7}
\end{figure}

\section{Gamma-ray fluxes}

The previous considerations determine the values of the proton injection
rate, the electron/proton ratio and the magnetic field strength $B_0$. The
only important parameter which cannot be determined from the analysis of
the synchrotron emission data is the external density $N_\mathrm{H}$.
Therefore we now compare calculations for different 
$N_\mathrm{H}$ with
the existing \gr measurements.

Figs.~\ref{fig7} and \ref{fig8} show the calculated integral 
$\gamma$-ray fluxes from the North-Eastern (NE) half of the remnant, in order 
to compare with the reported CANGAROO emission from the NE rim and the 
corresponding upper limit from H.E.S.S. These calculated fluxes correspond to 
50 percent of the entire theoretical \gr flux from SN~1006 under our assumption 
of dipolar symmetry \citep{vbk03}.

The key point is that a decrease of the ISM density leads to a considerable 
reduction of the expected $\pi^0$-decay \grs (see Fig.~\ref{fig7}): in 
the Sedov phase the $\pi^0$-decay \gr flux $F_{\gamma}^{\pi}\propto 
N_\mathrm{H} E_\mathrm{c}$ is proportional to the gas density and to the CR 
energy content $E_\mathrm{c}$. This roughly implies $F_{\gamma}^{\pi}\propto 
N_\mathrm{H}^2$, as shown in the previous section and in agreement with 
the numerical results. In addition, an increase of the gas density leads to an 
increase of the Alfv$\acute{e}$nic shock Mach number and of the shock 
modification due to 
CR backreaction. This implies in turn a harder 
CR spectrum and correspondingly an increase of the \gr flux at the 
highest energies. Only this last factor influences the IC \gr flux. Therefore 
it is only weakly sensitive to the gas density as Fig.~\ref{fig7} also 
indicates.

To allow for a simple reading of the figures, we 
present the same data in two
separate figures, Fig.~\ref{fig7} and Fig.~\ref{fig8}a, showing in
Fig.~\ref{fig7} the theoretical $\pi^0$-decay $F_{\gamma}^{\pi}$ and IC
$F_{\gamma}^\mathrm{IC}$ integral fluxes, whereas in Fig.~\ref{fig8}a we
show total fluxes $F_{\gamma}=F_{\gamma}^{\pi}+F_{\gamma}^\mathrm{IC}$ and
$F_{\gamma}^\mathrm{IC}$.

In Fig.~\ref{fig8} (see online version of this paper) we show the total ($\pi^0$-decay + IC) and the IC 
integral $\gamma$-ray fluxes, again from the NE half of the remnant as a 
function of $\gamma$-ray energy for the same cases as in Fig.~\ref{fig1}.
Also the total \gr flux varies strongly with the ISM density (Fig.\ref{fig7}).

A modification of the magnetic field strength would not 
change much the $\pi^0$-decay \gr flux, and thus the total \gr flux, whereas the 
IC flux decreases considerably with increasing magnetic field strength. 
Fig.~\ref{fig8}b illustrates this 
fact, even though we consider the value $B_0 \approx 30 \mu$G 
as experimentally quite well determined.
 
Note that an Inverse Compton \gr emission scenario in a low magnetic field 
of order $B_d = 4 {\pm 1} $~$\mu$G \citep{tanimori01} is clearly 
excluded by the H.E.S.S. upper limits, and this is fully consistent 
with the measured and expected field amplification.

The non-detection of SN~1006 by the H.E.S.S. experiment \citep{aharonian05} 
implies that the hydrogen density $N_\mathrm{H}$ is lower 
than $0.1$~cm$^{-3}$, and excludes the high value 
$N_\mathrm{H} = 0.3$~cm$^{-3}$ which happened to fit the CANGAROO data. As 
clearly shown in Fig.~\ref{fig8}a, at 1 TeV and for $B_0=30 \mu$G the 
H.E.S.S. upper limit is still about three times larger 
than the total \gr flux from SN~1006, expected for $N_\mathrm{H} = 
0.05$~cm$^{-3}$, which is at the lower end of the range $0.05 \leq N_\mathrm{H} 
\leq 0.3$~cm$^{-3}$ of plausible ambient densities.

Since the IC \gr emission is quite insensitive to the ISM density, the {\it
lower limit} for the expected \gr flux at TeV energies -- as derived from
the integrated synchrotron flux and the field amplification alone -- is about a
factor of five lower than the H.E.S.S. upper limit. Such a low \gr flux
from SN~1006 would be expected if the ambient ISM number density
$N_\mathrm{H}$ was even considerably lower than 0.05~cm$^{-3}$.

\section{Summary}

Our reexamination of the most relevant set of physical and astronomical 
parameters of SN~1006 led us to chose the somewhat greater
source distance of 
2.2 kpc, determined by more recent measurements, and to establish the interior 
magnetic field in SN~1006 more precisely at $B_\mathrm{d}\approx 150$~$\mu$G. 
Most important, however, is another fact: the lack of a TeV signal from SN~1006 
that follows from the non-detection by the H.E.S.S. instrument does not 
invalidate the theoretical picture which gives a consistent description of the 
nonthermal emission characteristics; it rather implies a constraint on the 
ambient gas density $N_\mathrm{H} < 0.1$~cm$^{-3}$. This point was 
discussed in detail. The renormalization of the 
\gr flux on which this result is partly based has been recently confirmed 
experimentally. The well-defined morphology makes this remnant the simplest and 
best understood of its kind. In conjunction with the amplification of the 
magnetic field by the accelerated cosmic rays, the H.E.S.S. non-detection 
completely rules out the low-field Inverse Compton scenarios that were 
popular until 
a few years ago.

In the case of SN~1006 and in other similar cases \gr measurements will 
at the same time give a reliable estimate of the ambient ISM density.

\begin{figure}[t]
 \includegraphics[width=.47\textwidth]{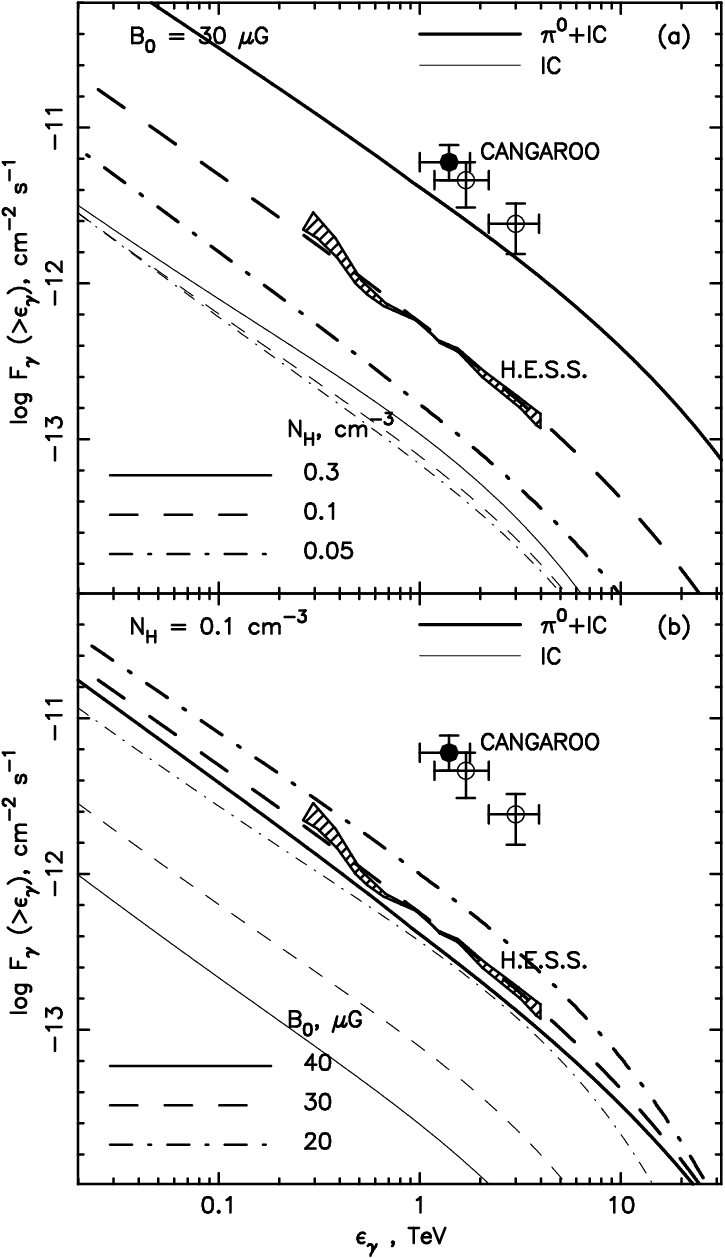}
 \caption{{\bf(On-line-only)} Total ($\pi^0$-decay + IC)
 ({\it thick lines}) and IC ({\it thin lines}) integral  
 $\gamma$-ray fluxes from the NE half of the remnant
 as a function of $\gamma$-ray energy for the same
 cases as in Fig.~\ref{fig1}. The CANGAROO-I ({\it open circles with error
 bars}) and CANGAROO-II ({\it solid circle with error bars}) flux from the NE 
rim
 \citep{hara02} and the corresponding H.E.S.S. upper 
limit for the CANGAROO position 
\citep{aharonian05} are shown as well. }
\label{fig8}
\end{figure}

Quantitatively (see Fig.~\ref{fig7} and Fig.~\ref{fig8}a), the present 
H.E.S.S. upper limit 
is only a factor of about three above the total TeV flux expected for 
$N_\mathrm{H} = 0.05$~cm$^{-3}$, which is at the lower end of the range $0.05 
\leq N_\mathrm{H} \leq 0.3$~cm$^{-3}$ discussed in the literature.
If the actual ISM is so diluted that $N_\mathrm{H} < 0.05$~cm$^{-3}$,
then the lowest expected TeV \gr flux is by a factor of five below the 
present H.E.S.S.
upper limit.

Given the arguments above, it is certainly worth attempting the 
detection of 
this important source in a deep observation of about 200h with the 
H.E.S.S. experiment.
%


\begin{acknowledgements} We thank the referee for a number of valuable 
suggestions to improve the paper. This work has been supported in part by the
Russian Foundation for Basic Research (grant 03-02-16524) and LSS 422.2003.2.
\end{acknowledgements}



\Online

\end{document}